\documentclass[apj]{emulateapj}
\usepackage{graphicx,epsf}
\newcommand{\ltsim}{\protect\raisebox{-0.5ex}{$\:\stackrel{\textstyle <}{\sim}\:$}}
\newcommand{\gtsim}{\protect\raisebox{-0.5ex}{$\:\stackrel{\textstyle >}{\sim}\:$}}

\shorttitle{Protostellar Collapse in Low-Metallicity Environments}
\shortauthors{K. Tomida}
\begin{document}
\title{Radiation Magnetohydrodynamic Simulations of Protostellar Collapse: Low-Metallicity Environments}

\author{Kengo Tomida}
\affiliation{Department of Astrophysical Sciences, Princeton University, Princeton, NJ 08544, USA; \mbox{tomida@astro.princeton.edu}}

\begin{abstract}
Among many physical processes involved in star formation, radiation transfer is one of the key processes since it dominantly controls the thermodynamics. Because metallicities control opacities, they are one of the important environmental parameters which affect star formation processes. In this work, I investigate protostellar collapse in solar-metallicity and low-metallicity ($Z = 0.1 Z_\odot$) environments using 3D radiation hydrodynamic and magnetohydrodynamic simulations. Because radiation cooling is more effective in the low-metallicity environments, first cores are colder and have lower entropies. As a result, first cores are smaller, less massive and have shorter lifetimes in the low-metallicity clouds. Therefore, first cores would be less likely to be found in low-metallicity star forming clouds. This also implies that first cores tend to be more gravitationally unstable and susceptible to fragmentation. The evolution and structure of protostellar cores formed after the second collapse weakly depend on metallicities in the spherical and magnetized models despite the large difference in the metallicities. Because this is due to the change of the heat capacity by dissociation and ionization of hydrogen, it is a general consequence of the second collapse as long as the effects of radiation cooling are not very large during the second collapse. On the other hand, the effects of different metallicities are more significant in the rotating models without magnetic fields, because they evolve slower than other models and therefore more affected by radiation cooling.
\end{abstract}


\section{Introduction}
Stars are formed in various environments with different density, temperature, rotation, turbulence, magnetic fields, metallicities, and so on. To take account of these effects, computational simulations including many physical processes have been extensively performed and played leading roles in understanding star formation processes. Among these physical processes, metallicities are one of the important parameters which control thermal evolution through radiation transfer. Thermodynamics is of critical importance in dynamics of collapsing clouds; for instance, the effective adiabatic index $\gamma$ is an important criterion whether a gas sphere collapses dynamically by self-gravity.

It is well accepted that the early phase of star formation, from molecular clouds cores to protostellar cores, proceeds through two distinctive stages \citep{lrs69,mi00,vaytet13}. Initially a dense cloud core collapses almost isothermally because radiation cooling is highly efficient. When the gas gets dense enough and radiation cooling becomes inefficient, the gas pressure balances with gravity and a quasi hydrostatic object, a so-called first core, is formed. This first core evolves by gas accretion from envelope. When the gas temperature around the center of the cloud reaches about 2,000K where molecular hydrogen starts to dissociate, the first core becomes unstable and collapses again. After this second collapse, finally a second core, or a protostellar core, is formed when the dissociation almost completes and the gas becomes stiff again. Modern multi-dimensional simulations with many physical processes including radiation transfer confirmed this scenario \citep{bate98,bate10,bate11,sch11,tomida13} and now it is well established. By the nature of its formation, it is obviously expected that the evolution and properties of first cores depend on radiation transfer and therefore on metallicities. However, it is not very clear how the formed protostars are affected by the different evolution in the early phase.

The effects of metallicities have been studied quite well, especially in the context of formation of the first and second generation stars. Using one-zone approximation or 1D hydrodynamic calculations, the effects of different metallicities on thermal evolution has been investigated elaborately \citep{omukai00,schn02,omukai05,ohy10}. Multi-dimensional hydrodynamic simulations with the barotropic approximation using the evolution track obtained in those calculations have been also performed, and conditions and outcomes of fragmentation have been investigated \citep{mcd08,mcd09}. On the other hand, in the context of present-day star formation, \citet{myers11} performed radiation hydrodynamic simulations of cluster-scale star formation, and found that the initial mass function is insensitive to the metallicities. While a statistical study on the large scale was done in their work, their resolution is limited ($\sim 7 {\rm AU}$) and optically-thick objects (i.e. first cores and circumstellar disks) are not well resolved, where the effects of different metallicities are expected to be more significant.

In this paper, I report the results of 3D nested-grid RHD and RMHD simulations of protostellar collapse from molecular cloud cores to protostellar cores with different metallicities. Because of the limited physical processes involved in the code, I focus on the $0.1 Z_\odot$ models and compare them with the solar abundance models in the context of present-day low-mass star formation. The main goal of this work is to clarify the effects of the metallicities especially on the structures and evolution of first cores and protostellar cores. This paper is organized as follows. In \S~2 the numerical methods and setups used in this paper are explained. The results of the simulations and analyses are shown in \S~3. Finally conclusions and discussions are presented in \S~4.

\section{Methods and Models}
I calculate collapse of unstable molecular cloud cores until protostellar cores are formed. The simulations are performed using the ${\rm ngr^3mhd}$ code which includes 3D nested-grid, resistive MHD, self-gravity, radiation transfer and a realistic equation-of-state (EOS). In this work, I do not use the resistive MHD part and I perform ideal MHD simulations and non-magnetized hydrodynamic simulations. The detailed description of the code is given in \citet{tomida13} (hereafter Paper I). The MHD part is solved by the HLLD approximate Riemann solver \citep{miyoshi} with second-order linear reconstruction. The solenoidal constraint is numerically enforced using the mixed correction proposed by \citet{dedner}. To solve the Poisson equation for self-gravity, I use the multi-grid solver \citep{mh03}. The gray Flux Limited Diffusion approximation (FLD; \citet{lp81,lev84}) is adopted for radiation transfer and the radiation subsystem is integrated implicitly. Finally, I use the tabulated equation-of-state which includes the effects of chemical reactions between seven species (${\rm H_2, H, H^+, He, He^+, He^{2+}}$ and ${\rm e^-}$) and internal degrees of freedom (rotation and vibration). The ortho:para ratio of molecular hydrogen is fixed to be 3:1.

To simulate low-metallicity clouds, the opacity tables are modified from Paper I. Although dust properties such as the structure, composition and size distribution can depend on the metallicities and other environmental parameters \citep[e.g.][]{singsdust,pldust}, I adopt the dust opacity tables provided by \citet{semenov} simply assuming that the dust opacities are proportional to the metallicities. For gas opacities, I combine the tables of \citet{fer05} and \citet{op94} with adequate abundances. In order to isolate the effects of different metallicities, other model parameters, the EOS in particular, are kept unchanged. Because the contribution from heavy elements is small, the effects of the different metallicities on the EOS are small. Note that clouds with $Z=0.1Z_\odot$ can be reasonably simulated with the current code but it is not adequate if the metallicity is below $Z\ltsim 10^{-2} Z_\odot$. In clouds with such low metallicities, radiation cooling by molecular lines is significant in the relatively low-density region \citep{ohy10}, but it is not implemented in the code as the dust continuum is assumed to be the dominant opacity source in the low-temperature region. Therefore, here I only study $Z=0.1 Z_{\odot}$ cases and compare them with the solar abundance models. I use $X=0.7$, $Z=0.02$ and $Y=1-(X+Z)$ for the solar abundance models, while the low metallicity models are calculated with $X=0.7$, $Z=0.002$.

I calculate six models; a spherical model, a rotating model without magnetic fields and a magnetized rotating model for each metallicity. The model parameters are summarized in Table.~\ref{tb:models}. The first character of the model corresponds to the initial conditions and the second represents the metallicities. In all the models, the Bonnor-Ebert spheres \citep{bonnor,ebert} with $20\%$ density enhancement are used as the initial conditions. The central density is $1.2\times 10^{-18}\, {\rm g\, cm^{-3}}$, the mass is $1 M_\odot$ and the radius is $\sim 8800 {\rm AU}$. Rigid-body rotation with $\Omega = 1.2\times 10^{-14} \, {\rm s^{-1}}$ is imposed for the rotating models as well as regularlized $10\% $ m=2 perturbation. Finally uniform mangetic fields of $ 20 \,{\rm \mu G}$ aligned to the rotational axis are introduced in the magnetized models. These initial conditions are similar to the slow rotating models in Paper I. The simulations are stopped when the central temperature reaches $T_c = 50,000 \, {\rm K}$ in the spherical and magnetized models, but the non-magnetized models are stopped at $T_c = 2,000 \, {\rm K}$ because they take much longer and I could not calculate the evolution after the second collapse long enough.

\begin{table}[tbp] 
\begin{center}
\begin{tabular}{c|cccccccc}
& $\Omega t_{\rm ff}$ & $\Omega \, (\times 10^{-14}\, {\rm s^{-1}})$ & $B_0 \, ({\rm\mu G})$ & $\mu_0$ & $A_2$ & $Z (Z_\odot)$ \\
\hline
{\it SS} & 0 & 0 & 0 & $\infty$ & 0 & 1\\
{\it SL} & 0 & 0 & 0 & $\infty$ & 0 & 0.1\\
{\it HS} & 0.023 & 1.2 & 0 & $\infty$ & 0.1 & 1\\
{\it HL} & 0.023 & 1.2 & 0 & $\infty$ & 0.1 & 0.1\\
{\it MS} & 0.023 & 1.2 & 20 & 3.8 & 0.1 & 1\\
{\it ML} & 0.023 & 1.2 & 20 & 3.8 & 0.1 & 0.1
\end{tabular}
\caption{}
{Summary of the initial model parameters. From left to right: the normalized angular velocities ($t_{\rm ff}\sim 6.08\times 10^{4} \, {\rm yrs}$), the angular velocities, the magnetic field strengths, the mass-to-flux ratios normalized to the critical value ($\mu_0\equiv \frac{M/\Phi}{(M/\Phi)_{\rm crit}}$ where $\Phi=\pi R^2 B_0$ and $(M/\Phi)_{\rm crit}=\frac{0.53}{3\pi}\left(\frac{5}{G}\right)^{1/2}$), the amplitudes of $m=2$ perturbation and the metallicity. Other parameters are common: $M=1\,M_\odot$, $R\sim 8800\, {\rm AU}$, $\rho_c=1.2\times 10^{-18}\, {\rm g\, cm^{-3}},$ and $T_0=10\,{\rm K}$.\label{tb:models}}
\end{center}
\end{table} 

\section{Results}
\subsection{Spherical Models}

\begin{figure*}[htb]
\begin{center}
\scalebox{0.3125}{\includegraphics{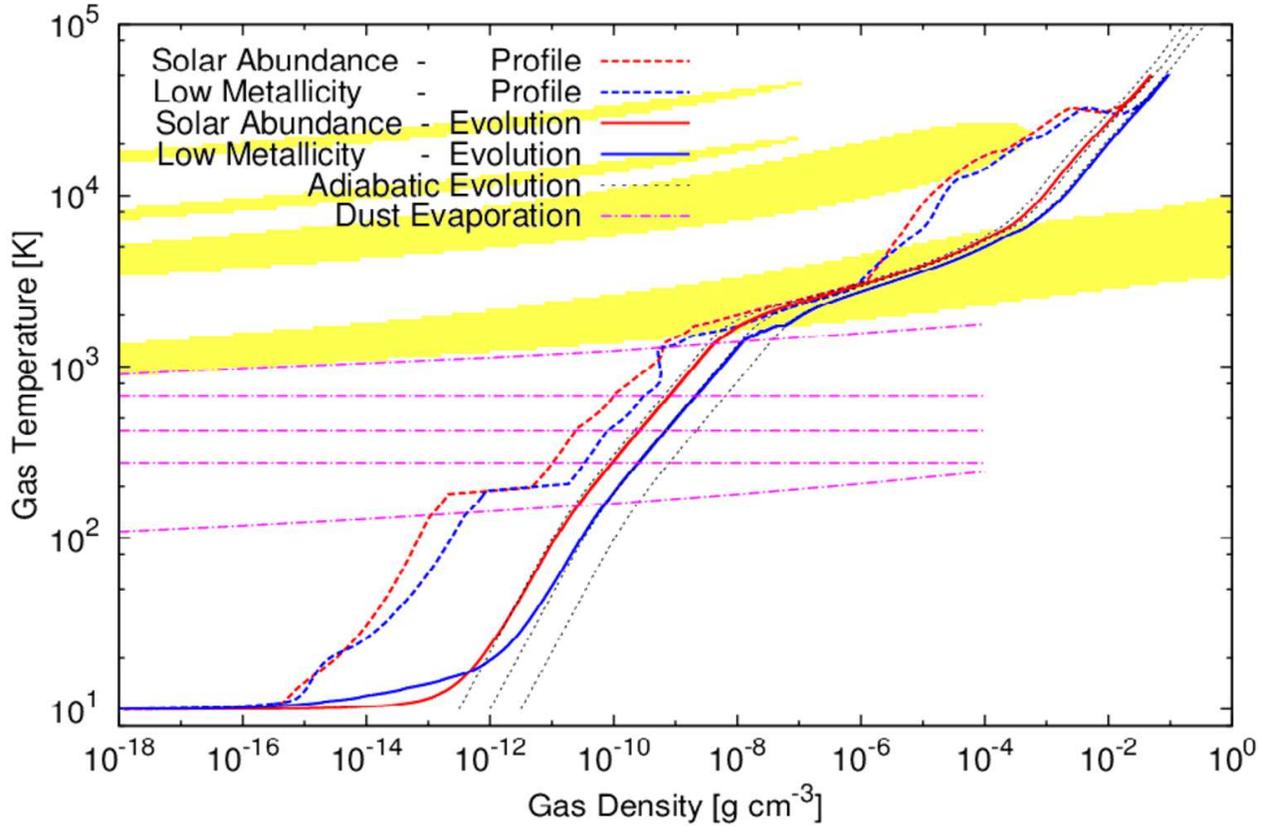}}
\caption{The profiles of the spherical models ({\it SS} and {\it SL}) at the end of the simulations (dashed lines) and the evolution tracks of the central gas elements (solid lines) in the $\rho-T$ plane. The red lines are the solar abundance model and the blue lines are the low-metallicity models. The gray dotted lines indicate purely adiabatic evolution tracks starting from $\log \rho = -11.5, -12.0, -12.5$ at $T=10\, {\rm K}$ and the magenta dash-dotted lines are the dust evaporation temperatures \citep{semenov}. The yellow shaded regions are where the adiabatic index $\gamma$ is below the critical value of the gravitational stability ($\gamma < 4/3$) due to the endothermic chemical reactions, which are dissociation of molecular hydrogen, ionization of hydrogen, the first and second ionizations of helium from bottom to top.}
\label{lm_rhot}
\end{center}
\end{figure*}

\begin{figure}[htb]
\begin{center}
\scalebox{0.89}{\includegraphics{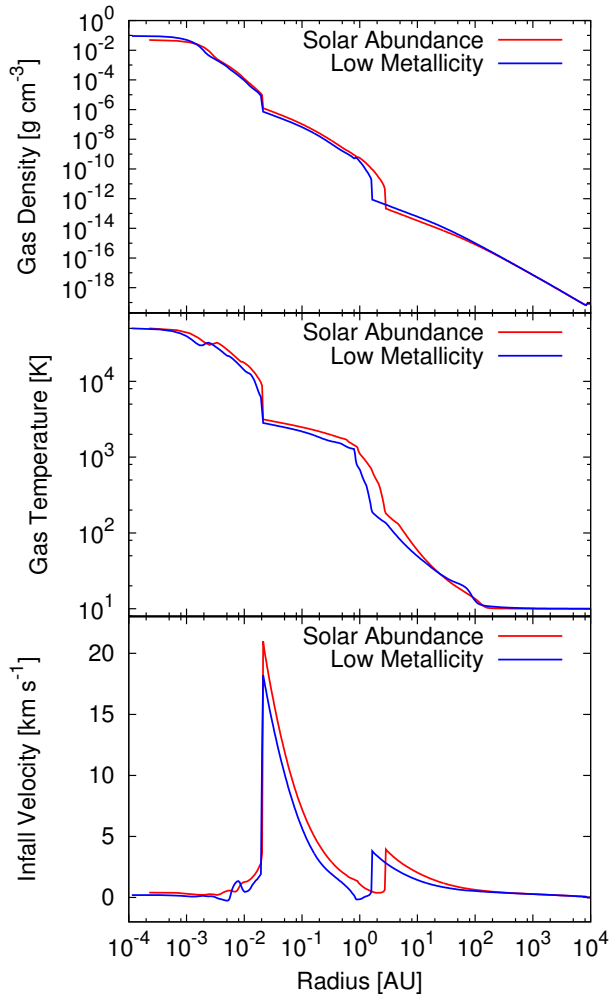}}
\caption{The radial distributions of the gas density, temperature and infall velocity (from top to bottom) in the spherical models at the end of the calculations. The red lines are the solar abundance model and the blue lines are the low-metallicity models.}
\label{lm_prof}
\end{center}
\end{figure}

\begin{figure}[htb]
\begin{center}
\scalebox{0.89}{\includegraphics{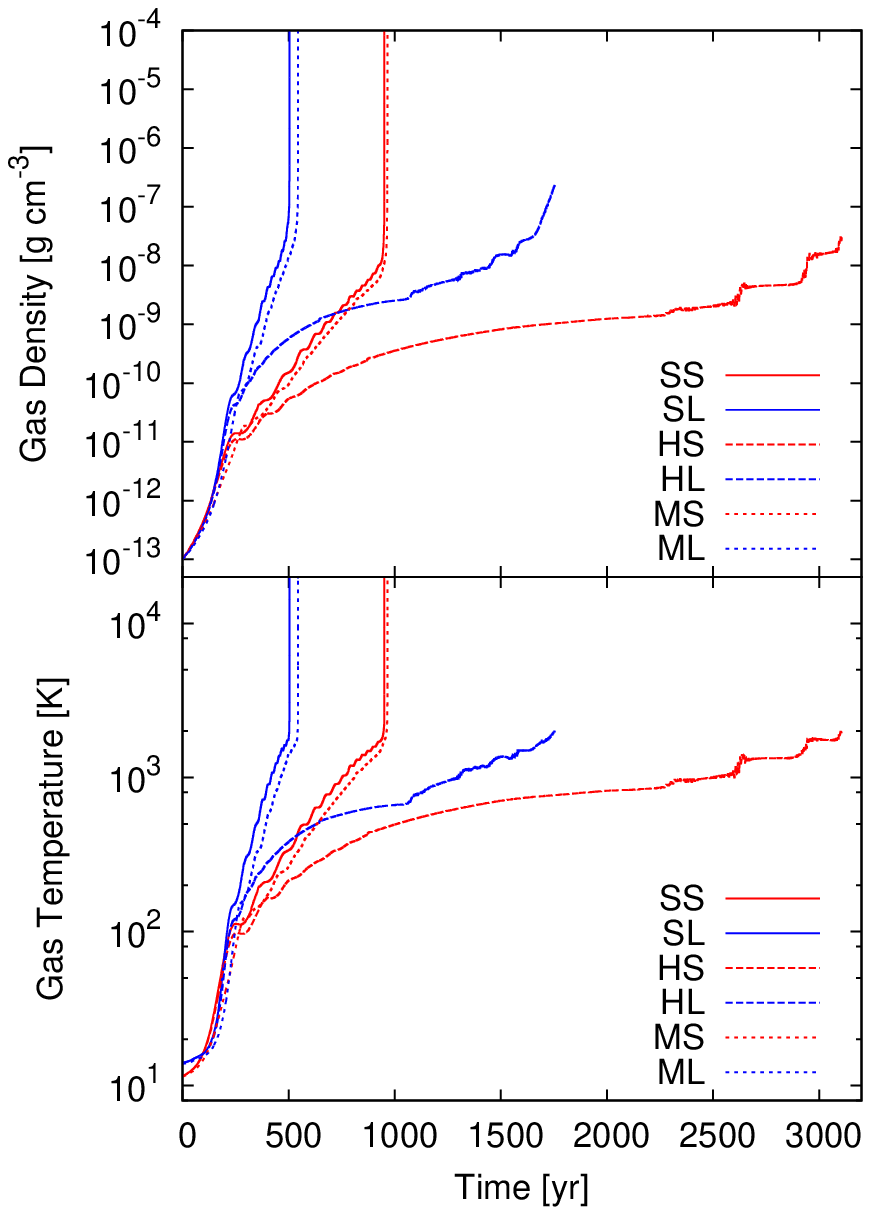}}
\caption{The density and temperature evolution at the center of the clouds as functions of the time. Colors are the same as in Fig.~\ref{lm_prof}. The spherical models are solid lines, the rotating models without magnetic fields are dashed, and the magnetized models are dotted lines. The origin of time is defined as the epoch when the central density reaches $10^{-13}\, {\rm g\, cm^{-3}}$.}
\label{lm_time}
\end{center}
\end{figure}

In order to clarify the effects of the different metallicities on the thermodynamics and dynamics, I discuss the spherical models first. Fig.~\ref{lm_rhot} shows the evolution tracks of the gas elements at the center of the clouds and the profiles at the end of the simulations in the $\rho-T$ plane. The radial density and temperature profiles at the end of the calculations are shown in Fig.~\ref{lm_prof}. Fig.~\ref{lm_time} is the time evolutions of the gas density and temperature after the first core formation.

\subsubsection{Overview of Evolution}
In the low-density regions before first cores are formed ($\rho_c\ltsim 10^{-12} \, {\rm g\, cm^{-3}}$), the gas evolves almost isothermally because of efficient radiation cooling. In this phase, the gas temperature rises earlier in the low-metallicity model. This is because radiation cooling is regulated by coupling between gas and radiation rather than radiation transfer in this phase, and its time scale is inversely proportional to the opacity. However, because the energy exchange is extremely sensitive to the gas temperature (note that the dust opacities are proportional to square of the temperature, $\kappa\propto T^2$, in the low temperature region), the gas temperature increases earlier but only slightly in the low metallicity case, and the gas collapses quasi-isothermally and dynamically until radiation transfer becomes inefficient.

When the gas at the cloud center becomes dense enough, $\rho_c\gtsim 10^{-12} \, {\rm g\, cm^{-3}}$, radiation cooling becomes inefficient and gas evolution becomes almost adiabatic following the EOS. Then, quasi-hydrostatic first cores are formed and they evolve via accretion from the envelope. The first core in {\it SL} is formed at the higher density and the gas evolution takes the path with a lower entropy (or temperature) because of more efficient radiation cooling. The density at the same temperature is lower in {\it SL} by a factor of $\sim 3$. Initially the adiabatic index is $\gamma \sim 5/3$ because the gas is too cold to excite rotational degrees of freedom of molecular hydrogen, but then it decreases to $\gamma \sim 7/5$ when the gas temperature reaches $T\sim 100\, {\rm K}$.

The adiabatic evolution continues until all the dust components evaporate at $T\sim 1,400\, {\rm K}$ (the top magenta line in Fig.~\ref{lm_rhot}), then radiation cooling becomes effective again and the evolution tracks go below the adiabatic curves (the gray dotted lines) for a short time. This is more prominent in {\it SL}. Beyond this point the opacities are dominated by molecules and atoms. Soon after the dust evaporation, dissociation of molecular hydrogen begins ($T\gtsim 1800 \, {\rm K}$, the yellow-shaded region). During this phase, the cores collapse dynamically because of the strongly endothermic reaction. As this collapse is so fast that it essentially happens in the free-fall time scale ($t_{\rm ff} \ltsim 1\, {\rm yr}$) and the gas density increases quickly, radiation cooling becomes ineffective and the evolution tracks follow the adiabatic curves again. Finally protostellar cores are formed when the dissociation almost completes and they evolve quasi-adiabatically. 

The deviation of the evolution tracks from the adiabatic curves after the dust evaporation indicate that the gas entropies are reduced by radiation cooling in both models, and it is more significant in {\it SL}. Therefore, the protostellar core in {\it SL} experiences more strong radiation cooling before its formation. However, the separation between the two evolution tracks get narrowed after the second collapse. This is solely due to the chemical reactions as seen in the adiabatic curves converging after the second collapse. This convergence means that protostellar cores are less sensitive to the conditions before the second collapse. The theoretical explanation of this process is given in Appendix.

 Overall, these spherical models are in good agreement with 1D spherical models of \citet{ohy10} in which more detailed chemistry and radiation transfer are considered. The evolutions in the low density region before formation of the first cores are different because of the different initial and boundary conditions, but the evolution and structures of the first cores and protostellar cores are consistent, including the effects of the dust evaporation.

\subsubsection{First Cores}
The first core structures are significantly different between the two models. The first core in {\it SL} has a significantly colder profile (Figures~\ref{lm_rhot} and \ref{lm_prof}). Because the colder gas provides less pressure support, the resulting first core is smaller, less massive and shorter-lived in the low-metallicity model. The radii and masses of the first core are 1.6 AU and $ 1.7  \times 10^{-2}M_\odot$ in {\it SL}, while they are 2.7 AU and $3.1 \times 10^{-2} M_\odot$ in {\it SS}. The surfaces of the first cores are identified as the sharp isothermal jumps around $\rho\sim 10^{-12}\, {\rm g \, cm^{-3}}$ in Fig.~\ref{lm_rhot}, which mean that almost all the kinetic energy in the inflow is radiated away to the upstream in both models \citep{com11a}. Defining the birth time of the first core as the first bounce ($T_c \sim 100\, {\rm K}$), the first core lifetimes are about 700 years in {\it SS} and about 300 years in {\it SL} (Fig.~\ref{lm_time}). So the mass and radius are smaller, and also the lifetime of the first core is significantly shorter by a factor of $\sim 2$ in the $0.1 Z_\odot$ case. Note that the mass is almost proportional to the lifetime because the evolutions in the isothermal phase are essentially the same and the accretion rates are similar in both models as predicted from the Larson-Penston self-similar solution \citep{lrs69,pen69}.

When all the dust components evaporate, the opacities are significantly decreased. This feature can be seen clearly both in the evolution tracks and the profiles (Figures \ref{lm_rhot} and \ref{lm_prof}), especially in {\it SL}, as a steep temperature drop around $\sim 1.2 \, {\rm AU}$ and $\rho \sim 6 \times 10^{-10}\, {\rm g \, cm^{-3}}$. Within this radius, the temperature gradient is shallower because the opacities are significantly decreased and radiation transfer works efficiently, but this region is confined by the optically-thick wall and therefore does not affect the envelope significantly in this early phase.

While the first core radius is smaller in {\it SL}, the surface temperature is similar to {\it SS}, which means the first core is fainter. This is because the first core is less massive at the end of its lifetime in {\it SL}. Therefore, it is very difficult to find a first core in low metallicity environments because of its lower luminosity and shorter lifetime.

\subsubsection{Protostellar Cores}
Compared to the first cores, the protostellar cores are more similar, as expected from the discussion above. I stop the simulations when the central temperature is $T_c\sim 50,000\,{\rm K}$\footnote{Although the simulation parameters are the same, this threshold is earlier than the end of simulations in Paper I, therefore these results should not be directly compared with those in Paper I.}, but the protostellar cores are still expanding as reported in Paper I \citep[see also][]{lrs69, wn80b}. At this epoch, the radii and masses of the protostellar cores are $4.2 R_\odot$ and $5.6\times 10^{-3} M_\odot$ in {\it SL}, $4.4 R_\odot$ and $7.2\times 10^{-3} M_\odot$ in {\it SS}. The protostellar core surfaces appear as the sharp adiabatic jumps around $\rho\sim 10^{-5}\, {\rm g \, cm^{-3}}$ in Fig.~\ref{lm_rhot}, indicating that the accretion flow is still highly optically-thick and radiation cooling is inefficient even in the low metallicity model. The accretion rates measured at the surfaces of the protostellar cores ($\dot{M}=4\pi r^2 \rho v_{\rm r}$) are $2.7 \times 10^{-3} M_\odot \, {\rm yr^{-1}}$ and $5.2 \times 10^{-3} M_\odot \, {\rm yr^{-1}}$ in {\it SL} and {\it SS}, respectively. This higher accretion rate in {\it SS} is due to the higher temperature when the second collapse begins as the accretion rate is $\dot{M}\sim c_s^3/G$. Thus, the metallicities affect the structure and early evolution of the protostellar cores but the differences are below a factor of 2 or less in spite of the metallicity difference of one order of magnitude. Note that these differences originate from the different conditions before the second collapse and radiation cooling after the second collapse is not important because the accretion flows are extremely optically thick. Thus, protostellar cores formed in different metallicity environments will be similar if other conditions like initial masses, turbulence and magnetic fields are similar.

\subsection{Rotating Models without Magnetic Fields}
In this section, I only discuss the first cores because the simulations are stopped when the central temperature reaches $T_c = 2,000 \,{\rm K}$ in the non-magnetized models. These simulations are much more computationally expensive than other models because they evolve much slower because of rotational support.

\begin{figure}[tb]
\begin{center}
\scalebox{0.68}{\includegraphics{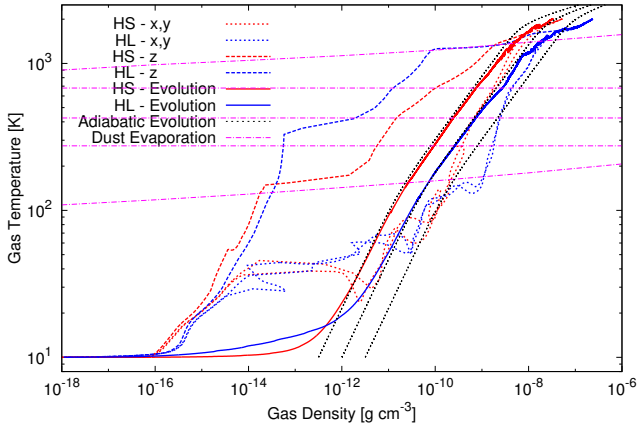}}
\caption{The profiles of the non-magnetized rotating models ({\it HS} and {\it HL}) at the end of the simulations along the rotation axis ($z$, dashed lines) and in the disk mid-planes ($x$ and $y$, dotted lines),  and the evolution tracks of the central gas elements (solid lines) in the $\rho-T$ plane. The red lines are the solar abundance model and the blue lines are the low-metallicity models. The adiabatic evolution tracks and dust evaporation temperatures are also plotted as in Fig.~\ref{lm_rhot}. Note that the density and temperature ranges are different from Figures \ref{lm_rhot} and \ref{lm_mhds}.}
\label{lm_rot}
\end{center}
\end{figure}

\begin{figure*}[t]
\begin{center}
\scalebox{0.45}{\includegraphics{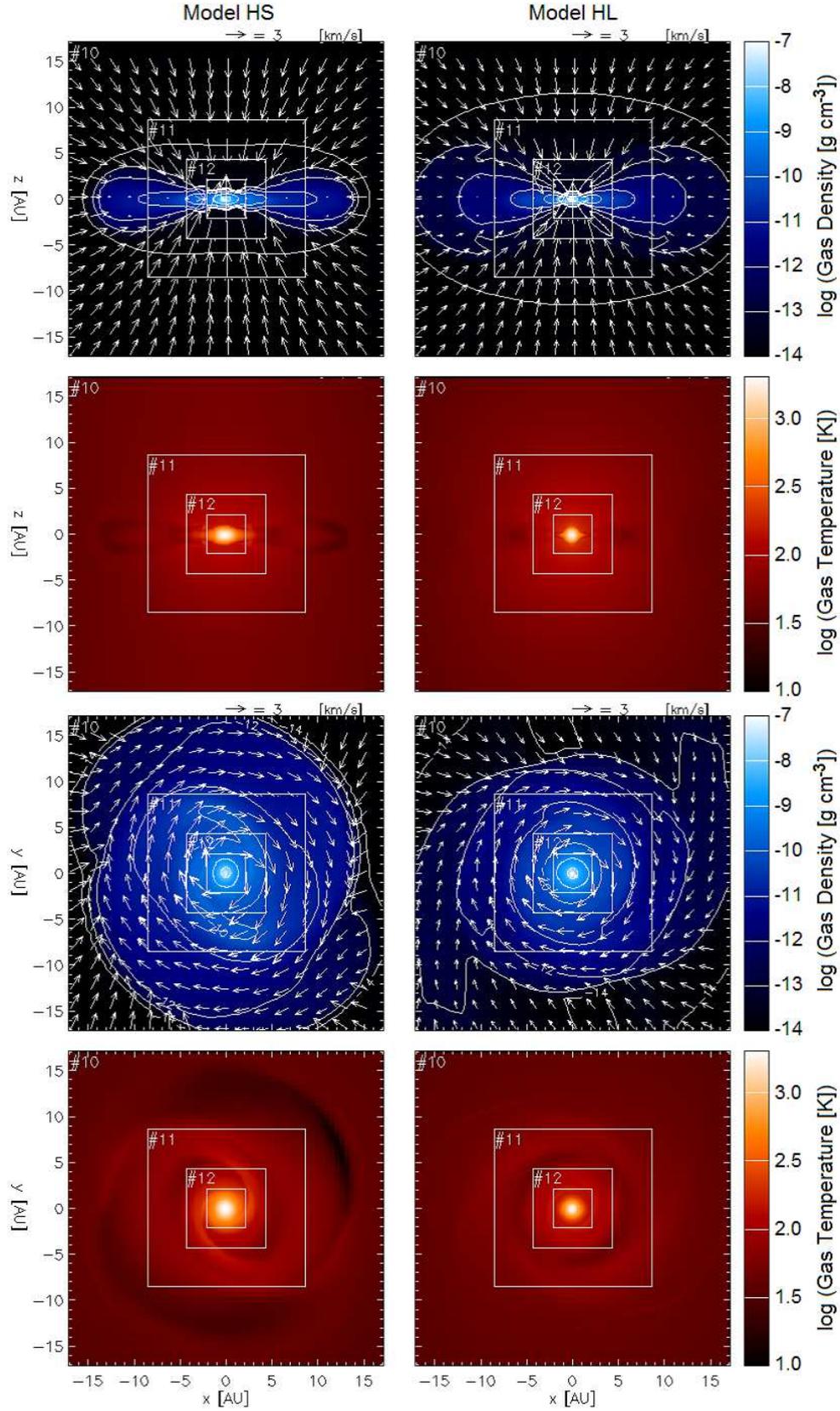}}
\caption{The cross sections of the rotation models without magnetic fields (the left column is {\it HS} and the right is {\it HL}) at the end of the first core phases ($T_c\sim 2,000\, {\rm K}$).  The gas density is plotted in the first and third rows, while the others are temperature. The top two rows are edge-on view and bottom two are face-on view. The edge of each panel is about 35 AU.}
\label{rot_cs}
\end{center}
\end{figure*}

\subsubsection{Overview of Evolution}
The angular momentum within the initial molecular cloud core is large enough to prevent gas from directly collapsing into a single star. As a result, the first cores are supported by rotation and form circumstellar disks. As there are no magnetic fields in these models, these disk are so massive that they become gravitationally unstable and spiral arms are spontaneously formed. Then gravitational torque through these non-axisymmetric structures transport angular momentum efficiently, which enables formation of the protostellar cores \citep{bate98}.

The evolution tracks of the central gas elements (Fig.~\ref{lm_rot}) clearly show the significant effects of radiation cooling. Although the thermal evolutions in the early phase are similar to those in the spherical models, the evolution tracks deviate from the adiabatic evolution earlier and more significantly. This is obviously because the first cores evolve more slowly due to the rotational support (Fig.~\ref{lm_time}). The lifetimes of the first cores are about 2800 years in {\it HS} and about 1500 years in {\it HL}. Since the radiation cooling is more prominent in {\it HL}, the difference in the formed protostellar cores will be larger than in the spherical models.

\subsubsection{First Cores}
 The cross sections of the density and temperature at the end of the simulations are shown in Fig.~\ref{rot_cs}. The first cores are supported by rotation and form disks as the spiral arms are formed by the gravitational instability and transport angular momentum. The profiles in the disks (the dotted lines in Fig.~\ref{lm_rot}) show that the first core disks are strongly cooled by radiation, and they are even colder than the central gas elements. The effect of radiation cooling is more prominent in {\it HL} in the high density region ($\rho \gtsim 10^{-10} \, {\rm g\, cm^{-3}}$), while the low density regions have similar temperatures. The (major, minor) radii of the rotationally supported regions (where the toroidal velocity is dominant) and the masses of the first core are about (16 AU, 13 AU) and $ 5.6  \times 10^{-2}M_\odot$ {\it HL}, while they are (17 AU, 14 AU) and $8.8 \times 10^{-2} M_\odot$ in {\it HS}. Note that the radii are difficult to measure because they are highly dynamical as the gravitational instability occurs intermittently. While their masses differ by about 60\%, the sizes the rotationally-supported disks are similar, which means that the angular momentum transport by the gravitational torque is more efficient in the low metallicity environment because the disk is colder and therefore more gravitationally unstable. Although the disks in this work have not fragmented during the first core phase, these results suggest that the circumstellar disks tend to be more susceptible to fragmentation in the lower metallicity environments. 

The vertical structures at the disk centers (the dashed lines in Fig.\ref{lm_rot}) are significantly different. The thicknesses at the centers of the first cores are 1.1 AU in {\it HS} and 0.8 AU in {\it HL}, which are smaller than the radii in the spherical models. As the evolution in the rotating models is significantly slower than that in the spherical models, the temperature distributions do not have jumps any more at the dust evaporation temperature in both models. The dust evaporation front extends to the lower density region in {\it HL} and the surface temperature is significantly higher than in {\it HS}.
 
 The non-magnetized model with the solar abundance can be compared with \citet{bate11}. Although the overall evolutions are similar, the disk in {\it HS} is colder than $\beta=0.005$ model in \citet{bate11}, probably because of the different accretion rates due to the different initial temperatures (10K vs 14K) and density profiles (Bonnor-Ebert vs uniform). The accretion rate achieved in this work is about half of that in \cite{bate11} (estimated from the lifetimes of the first cores without rotation), which results in longer cooling and less heating. Also, since the opacities are proportional to $T^2$ in the low temperature, the higher initial temperature means the higher optical depth in the envelope.

\subsection{Rotating Models with Magnetic Fields}

\begin{figure}[tb]
\begin{center}
\scalebox{0.68}{\includegraphics{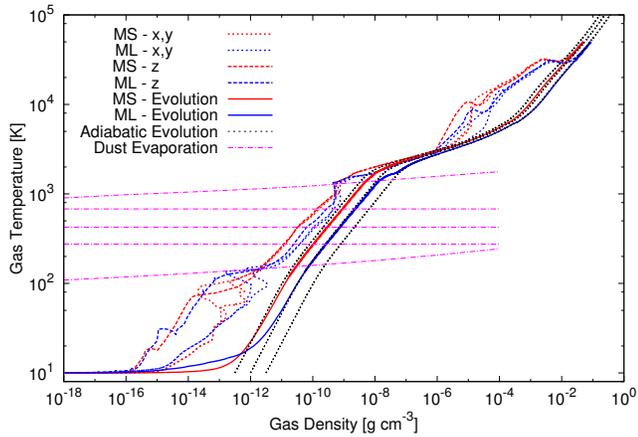}}
\caption{Same as Fig.~\ref{lm_rot} but of the magnetized models ({\it MS} and {\it ML}).}
\label{lm_mhds}
\end{center}
\end{figure}

\subsubsection{Overview of Evolution}
When magnetic fields present, they transport angular momentum very efficiently so that circumstellar disks are not formed in the early phase of star formation. This is the so-called magnetic braking catastrophe as discussed in many precedent works \citep[e.g.][]{ml08,ht08,li11}. As the magnetic fields interact with rotating gas, the bipolar outflows are launched around the first cores by the magneto-centrifugal force \citep{bp82,tmsk98,tmsk02,com10,tomida10a}, which also carry angular momentum away from the first cores. As a result, the thermal evolutions are very similar to those in the spherical models, which are clearly seen in Fig.~\ref{lm_mhds}. The lifetimes of the first cores are 700 years in {\it MS} and 300 years in {\it ML}, respectively, which are also essentially the same as in the spherical models (Fig.~\ref{lm_time}).

\begin{figure*}[tb]
\begin{center}
\scalebox{0.45}{\includegraphics{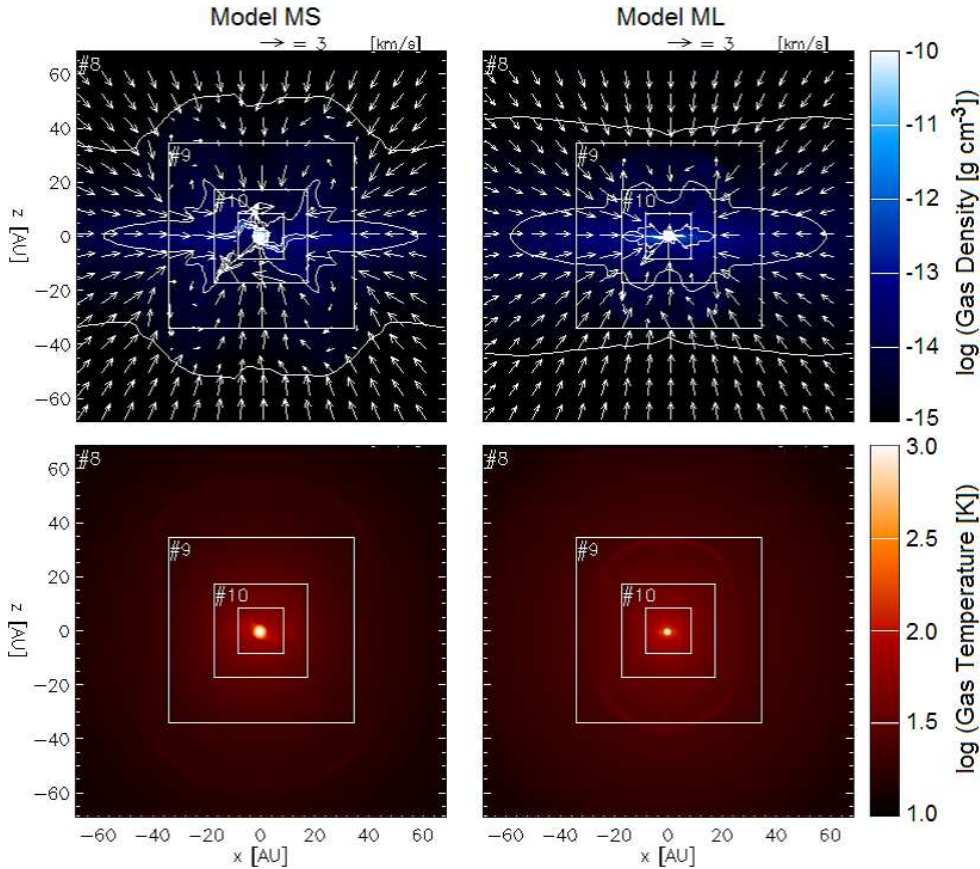}}
\caption{The density (top) and temperature (bottom) cross sections of the magnetized models (the left column is {\it MS} and the right is {\it ML}) in the outflow scale (the edge is about 140 AU).  Because the system is quasi-axisymmetric except for the perturbation due to the interchange instability, only the edge-on figures are shown.}
\label{of_cs}
\end{center}
\end{figure*}
\begin{figure*}[tb]
\begin{center}
\scalebox{0.45}{\includegraphics{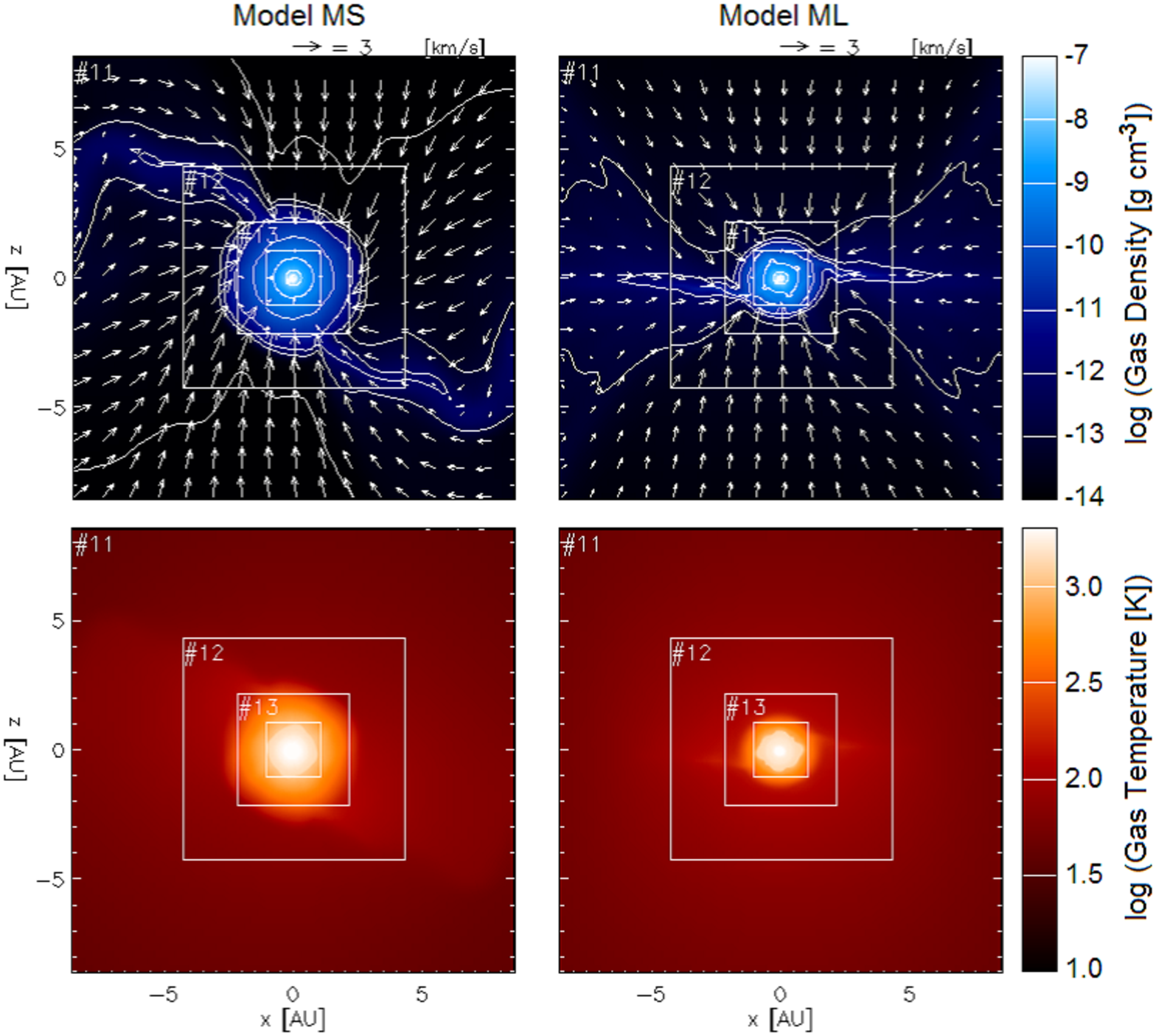}}
\caption{The same as Fig.~\ref{of_cs} but in the first core scale (the edge is about 18 AU).}
\label{fc_cs}
\end{center}
\end{figure*}

\subsubsection{First Cores and Outflows}
The density and temperature cross sections at the outflow scale and the first core scale are shown in Figures \ref{of_cs} and \ref{fc_cs}. Slow ($v\ltsim 1 \,{\rm km\ s^{-1}}$) molecular outflows with wide-opening angles are driven around the first cores in both models. Although the sizes of the outflows are different, their velocities and morphologies are very similar because the gas thermodynamics has almost no impact on the launching mechanism of the outflows. Therefore, roughly speaking, the different traveled distances of the outflows at the end of the simulations, about 55 AU in {\it MS} and 35 AU in {\it ML}, are simply consequences of the first core lifetimes.

The first cores and surrounding pseudo disks warp due to the magnetic interchange instability (see Paper I). Except for these perturbations, the first cores are virtually spherical in both models; the first cores are rotating only slowly as a result of the efficient angular momentum transport by magnetic fields. The distributions in Fig.~\ref{lm_mhds} also indicate that the first cores are spherically symmetric; all the profiles along three axes are similar in the first cores ($10^{-11}\,{\rm g\, cm^{-3}}\ltsim \rho \ltsim 10^{-8}\,{\rm g\, cm^{-3}}$). On the other hand, the outer regions ($\rho\ltsim 10^{-12}\, {\rm g\, cm^{-3}}$) are significantly anisotropic because there present the pseudo disks and outflows. The first core mass and radius in {\it MS} are $5.6\times 10^{-3} M_\odot$ and $3.0\, {\rm AU}$, while they are $1.9 \times 10^{-3} M_\odot$ and $1.6\, {\rm AU}$ in {\it ML}. These properties are again very close to those in the spherical models.

\begin{figure*}[tb]
\begin{center}
\scalebox{0.45}{\includegraphics{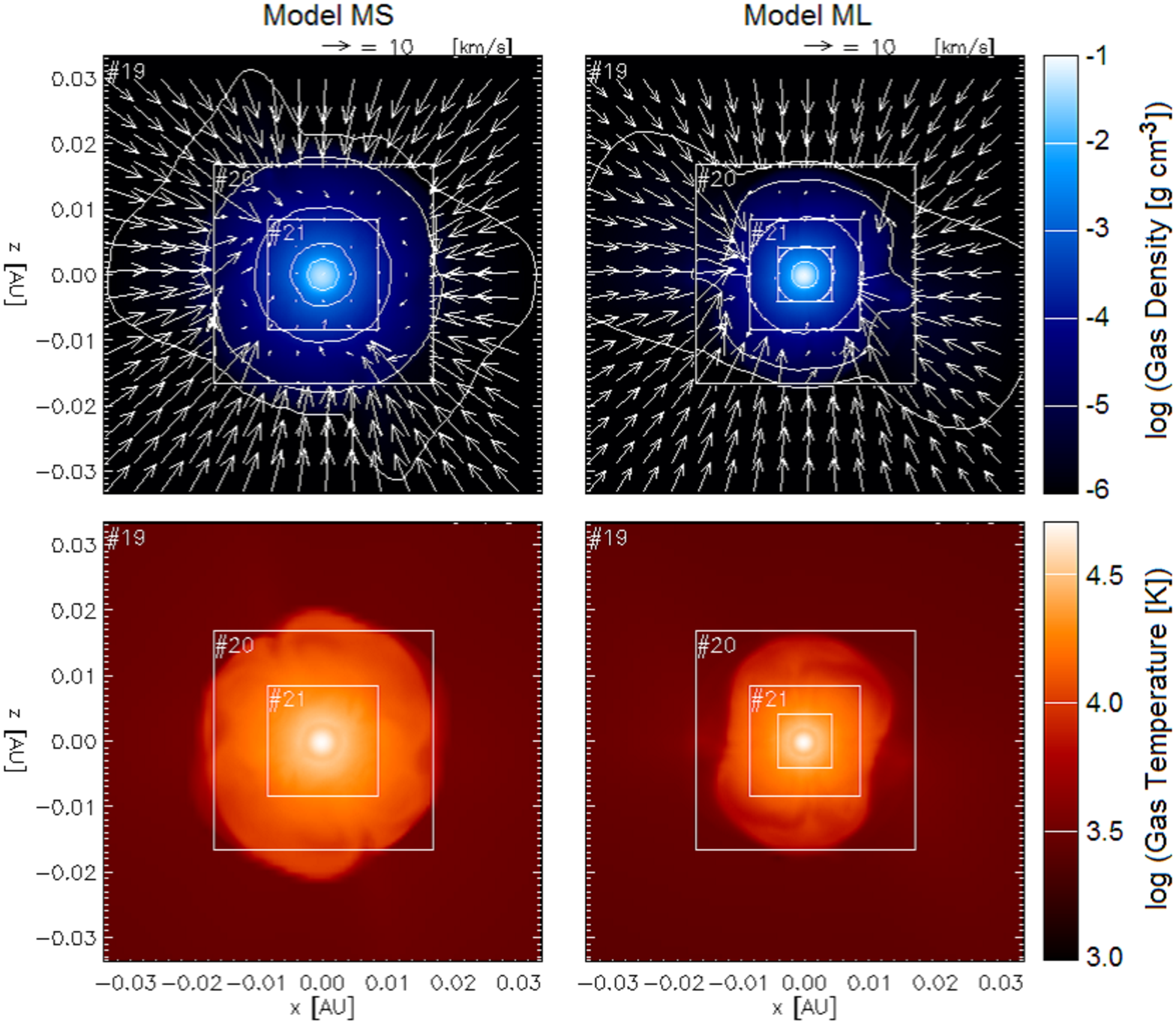}}
\caption{The same as Fig.~\ref{of_cs} but in the protostellar core scale (the edge is about 0.07 AU or 17 $R_\odot$).}
\label{pc_cs}
\end{center}
\end{figure*}
\subsubsection{Protostellar Cores}
The formed protostellar cores are rotating very slowly in both models, and not supported by rotation at all. They are slightly elongated along the rotational axis and the accretion flows are somewhat anisotropic because of the interchange instability in the first core scales, but roughly speaking the protostellar cores are almost spherically symmetric (Fig.~\ref{pc_cs}). In Fig.~\ref{lm_mhds}, the outer regions of the protostellar cores close to their surfaces ($\rho \sim 10^{-5}\, {\rm g\, cm^{-3}}$) have different profiles due to the anisotropy, but the central regions ($\rho \gtsim 10^{-4}\, {\rm g\, cm^{-3}}$) are highly symmetric. Their properties are essentially the same as in the spherical models; the radii and masses of the protostellar cores are $4.4 R_\odot$ and $6.9\times 10^{-3} M_\odot$ in {\it MS}, $3.9 R_\odot$ and $4.8 \times 10^{-3} M_\odot$ in {\it ML}. As in the spherical models, they are still expanding and evolving as the gas accretion continues. The protostellar core in {\it ML} is less massive and less evolved than in {\it MS} and also than in {\it SL} by some reason although the central density and temperature are almost the same as in {\it SL}. Again, the differences in the protostellar core properties are not as significant as the metallicity difference, which is one order of magnitude.

\section{Conclusions and Discussions}

\begin{table*}[tbp] 
\begin{center}
\begin{tabular}{c|cccccc}
Model & $M_{\rm FC} (10^{-2} M_\odot)$ & $R_{\rm FC}({\rm AU})$ & $t_{\rm FC} ({\rm yrs})$ & $M_{\rm PC}(10^{-3} M_\odot)$ & $R_{\rm PC}(R_\odot)$ & $L_{\rm out} ({\rm AU})$ \\
\hline
{\it SS} & 3.1 & 2.7 & 700 & 7.2 & 4.4 & -\\
{\it SL} & 1.7 & 1.6 & 300 & 5.6 & 4.2 & -\\
{\it HS} & 8.8 & 17 & 2800 & - & - & -\\
{\it HL} & 5.6 & 16 & 1500 & - & - & -\\
{\it MS} & 3.1 & 3.0 & 700 & 6.9 & 4.4 & 55\\
{\it ML} & 1.9 & 1.6 & 300 & 4.8 & 3.9 & 35
\end{tabular}
\end{center}
\caption{}{The properties of the first cores, protostellar cores and outflows in the simulations. From left to right, the masses, radii and lifetimes of the first cores, the masses, radii of the protostellar cores, and the sizes of the outflows. The largest radii are shown when the first and protostellar cores are not spherical. Note that the first and protostellar cores are still evolving and their properties are transient.\label{tb:restable}}
\end{table*} 

In this paper, I performed 3D RHD and RMHD simulations of protostellar collapse in low metallicity environments ($Z= 0.1 Z_\odot$) and compared them with the models with the solar metallicity. Note that the evolutions after dust evaporation in the low metallicity models are largely changed from the presentation at Protostars \& Planets VI because there were errors in the low-metallicity gas opacities. In order to simulate thermodynamics and dynamics in star formation processes, multi-dimensional simulations including radiation transfer are crucial because the thermal evolution is not a simple function of the local gas properties but highly non-linear, non-local and time-dependent. In particular, the barotropic approximation used in many previous works cannot properly reproduce the thermal properties.

The properties of the first cores, protostellar cores and outflows obtained in the simulations are summarized in Table.~\ref{tb:restable}. In short, the early evolution and structures of the protostellar cores weakly depend on the metallicities while the first cores are significantly more sensitive to the metallicities. Although I simulated only the earliest evolution of the protostellar cores, these results are useful as input for further stellar evolution studies. The detailed findings in this work are summarized as follows:

1. The first cores in the low-metallicity environments are colder because of more efficient radiation cooling. As a result, the low-metal first cores have smaller radii, masses and shorter lifetimes.

2. Therefore, first cores are less likely to be found in low-metallicity environments.

3. The first cores and protostellar cores in the ideal MHD models are very similar to those in the spherical models, except for the outflows. This is because of the efficient angular momentum transport by magnetic fields.

3. The molecular outflows in the MHD models are very similar in their speeds and morphologies because the thermodynamics does not affect the driving mechanism of the outflows. The different sizes of the outflows at the end of the first core phase are simply due to their lifetimes.

4. In the spherical and the ideal MHD models, the properties of the formed protostellar cores are not very sensitive to the metallicities. This is solely a consequence of the dissociation and ionization of hydrogen and therefore quite general (see Appendix for the theoretical explanation).  Since typical magnetic fields observed in star forming clouds are strong enough (e.g. \citet{crutcher} and references therein), the metallicity-dependencies of the protostellar cores are expected to be small at least when they are very young, while the first cores are more sensitive to the metallicities.

5. When there is rotation but no or very weak magnetic fields, the first cores form disks supported by rotation. As they evolve more slowly, the disks become significantly colder than the first cores in the spherical models due to radiation cooling. The effects of radiation cooling is more prominent in the low metallicity environments. As a result, considerable differences may appear in the protostellar cores compared to the spherical and magnetized cases. Although the first cores in this work have not fragmented, these results indicate that the disks would be more susceptible to the gravitational instability and fragmentation in the low-metallicity environments. Naively speaking, this implies a higher binary rate and probability of planetary-mass companions, which is qualitatively consistent with the previous simulations using the barotropic approximation \citep{mcd08,mcd09}. Note that, however, this is not trivial because the initial conditions can depend on the metallicities.

In order to predict the outcomes of star formation processes such as the initial mass function and the binary or multiple rates, it is crucial to convolve the initial distributions and the outcome of each formation process. As these results are just case studies on the early phases of the specific models, more broad parameter survey and long-term calculations are obviously required. For example, the initial rotation and magnetic fields are very important because the angular momentum and its transport are crucial in the evolution of the disks. Besides, the mass of the initial molecular cloud also has significant impact. The radiation cooling effects would be more significant in the low mass clouds because the evolution is slower due to the lower accretion rate \citep{tomida10b}. Although current observations of low metallicity star forming regions are limited, future observations of star forming clouds in the outer regions of our Galaxy or nearby satellites like SMC and LMC will be useful to understand the effects of metallicities.

As a future work, the effects of metallicities on non-ideal MHD effects are important and interesting. While the Ohmic dissipation extends the first core lifetimes slightly by suppressing the angular momentum transport (Paper I), the ionization degree and resulting resistivity themselves depend on the metallicities. In the relatively low-density region where cosmic rays predominantly contribute to ionization, the ionization degree would be higher and the resistivity would be lower than in the solar metallicity environments. On the other hand, in the relatively high density region, typically the interior of first cores, decay of radioactive nuclei is the dominant ionization source. Therefore, the ionization degree could decrease and the non-ideal MHD could be more effective. Also, as the abundance of potassium is smaller, the re-coupling due to the thermal ionization of potassium (K) would occur at a slightly higher density. Thus, the non-ideal MHD effects are not trivial and detailed studies are required.\\

The author thanks Prof. Eve Ostriker, Prof. James Stone, Prof. Kazuyuki Omukai, Prof. Masahiro Machida, Dr. Takashi Hosokawa and Dr. Shintaro Takayoshi for fruitful discussions and comments. Numerical computations were partly performed on NEC SX-9 at Center for Computational Astrophysics of National Astronomical Observatory of Japan and at Japan Aerospace Exploration Agency. KT is supported by Japan Society for the Promotion of Science (JSPS) Postdoctoral Fellowship for Research Abroad.

\section*{Appendix: Convergence of the Evolutions After the Second Collapse}
In this appendix, I explain why the adiabatic curves with different entropies converge after the dissociation of molecular hydrogen (Fig.~\ref{lm_rhot}). For simplicity, I only discuss hydrogen gas as contribution of helium is small.

Suppose two adiabatic tracks with the same number of hydrogen atoms but different entropies contracting from fully molecular gas to fully ionized gas. Track A starts from the initial state $A_1$ ($\rho_1, T^A_1$) and ends at the final state $A_2$ ($\rho_2, T^A_2$), while Track B starts from $B_1$ ($\rho_1, T^B_1$) and ends at $B_2$ ($\rho_2, T^B_2$). The entropy difference between the tracks can be estimated by considering isochoric processes $B_1 \rightarrow A_1$ and $B_2 \rightarrow A_2$:
\begin{eqnarray}
\Delta S = \int_{B}^{A} dS = \int_{B}^{A}C_V\frac{dT}{T}.
\end{eqnarray}
For simplicity, let us assume the heat capacity is constant during these processes and neglect non-ideal effects such as chemical reactions and interaction between particles. Then, the entropy difference is estimated to be
\begin{eqnarray}
\Delta S \sim C_V \ln \frac{T^A}{T^B}.
\end{eqnarray}
This indicates that the temperature difference is smaller when $C_V$ is larger, and it is most significant during the second collapse because dissociation of molecular hydrogen involves a huge latent heat and increases $C_V$. At $A_1$ and $B_1$ where the temperatures are so low ($T^{A,B}_1 < 100\, {\rm K}$) that rotational and vibrational degrees of freedom are not excited, the heat capacity is $C_V\sim \frac{3}{2}N_1k$, where N is the number of particles. At $A_2$ and $B_2$ where hydrogen is completely dissociated and ionized ($T^{A,B}_2 > 10^5\, {\rm K}$), the heat capacity is also $C_V\sim \frac{3}{2}N_2k$. Because the two tracks are adiabatic, the entropy difference is constant everywhere. Therefore a simple relation between the temperature ratios is derived:
\begin{eqnarray}
N_1 \ln \frac{T^A_1}{T^B_1}\sim N_2 \ln \frac{T^A_2}{T^B_2}.
\end{eqnarray}
As the gas is fully molecular in the initial states and fully ionized in the final states, $N_2 \sim 4 N_1$ (it is possible to improve this estimate by considering contribution from helium). Finally, 
\begin{eqnarray}
\frac{T^A_2}{T^B_2}\sim\left(\frac{T^A_1}{T^B_1}\right)^{\frac{1}{4}}.
\end{eqnarray}

Therefore, the temperature ratio after the second collapse is much smaller than that before the second collapse. The temperature ratio between the two tracks in Fig.~\ref{lm_rhot} starting from $\log \rho=-11.5$ and $-12.5$ is $T^A_1/T^B_1\sim 4.64$ at $\log \rho_1=-11.5$, while it is $T_{A_2}/T_{B_2}\sim 1.56$ at $\log \rho_2=0$. This is in good agreement with the theoretical prediction, $(4.64)^{1/4} \sim 1.47$, but slightly deviates from the prediction probably because of the neglected effects including contribution from helium and its ionization.

Because this significant effect is solely a consequence of thermodynamics and chemical reactions, it leads to a quite general prediction; as long as radiative loss during the second collapse is small (although this might not be the case in the non-magnetized rotating models discussed in the text), formed protostellar cores only weakly depend on the initial conditions.


\begin{thebibliography}{40}
\expandafter\ifx\csname natexlab\endcsname\relax\def\natexlab#1{#1}\fi

\bibitem[{{Bate}(1998)}]{bate98}
{Bate}, M.~R. 1998, \apjl, 508, L95

\bibitem[{{Bate}(2010)}]{bate10}
---. 2010, \mnras, 404, L79

\bibitem[{{Bate}(2011)}]{bate11}
---. 2011, \mnras, 417, 2036

\bibitem[{{Blandford} \& {Payne}(1982)}]{bp82}
{Blandford}, R.~D., \& {Payne}, D.~G. 1982, \mnras, 199, 883

\bibitem[{{Bonnor}(1956)}]{bonnor}
{Bonnor}, W.~B. 1956, \mnras, 116, 351

\bibitem[{{Commer{\c c}on} {et~al.}(2011){Commer{\c c}on}, {Audit}, {Chabrier},
  \& {Chi{\`e}ze}}]{com11a}
{Commer{\c c}on}, B., {Audit}, E., {Chabrier}, G., \& {Chi{\`e}ze}, J.-P. 2011,
  \aap, 530, A13

\bibitem[{{Commer{\c c}on} {et~al.}(2010){Commer{\c c}on}, {Hennebelle},
  {Audit}, {Chabrier}, \& {Teyssier}}]{com10}
{Commer{\c c}on}, B., {Hennebelle}, P., {Audit}, E., {Chabrier}, G., \&
  {Teyssier}, R. 2010, \aap, 510, L3+

\bibitem[{{Crutcher}(2012)}]{crutcher}
{Crutcher}, R.~M. 2012, \araa, 50, 29

\bibitem[{{Dedner} {et~al.}(2002){Dedner}, {Kemm}, {Kr{\"o}ner}, {Munz},
  {Schnitzer}, \& {Wesenberg}}]{dedner}
{Dedner}, A., {Kemm}, F., {Kr{\"o}ner}, D., {Munz}, C.-D., {Schnitzer}, T., \&
  {Wesenberg}, M. 2002, Journal of Computational Physics, 175, 645

\bibitem[{{Draine} {et~al.}(2007){Draine}, {Dale}, {Bendo}, {Gordon}, {Smith},
  {Armus}, {Engelbracht}, {Helou}, {Kennicutt}, {Li}, {Roussel}, {Walter},
  {Calzetti}, {Moustakas}, {Murphy}, {Rieke}, {Bot}, {Hollenbach}, {Sheth}, \&
  {Teplitz}}]{singsdust}
{Draine}, B.~T., {et~al.} 2007, \apj, 663, 866

\bibitem[{{Ebert}(1955)}]{ebert}
{Ebert}, R. 1955, Zeitschrift fur Astrophysik, 36, 222

\bibitem[{{Ferguson} {et~al.}(2005){Ferguson}, {Alexander}, {Allard}, {Barman},
  {Bodnarik}, {Hauschildt}, {Heffner-Wong}, \& {Tamanai}}]{fer05}
{Ferguson}, J.~W., {Alexander}, D.~R., {Allard}, F., {Barman}, T., {Bodnarik},
  J.~G., {Hauschildt}, P.~H., {Heffner-Wong}, A., \& {Tamanai}, A. 2005, \apj,
  623, 585

\bibitem[{{Hennebelle} \& {Teyssier}(2008)}]{ht08}
{Hennebelle}, P., \& {Teyssier}, R. 2008, \aap, 477, 25

\bibitem[{{Larson}(1969)}]{lrs69}
{Larson}, R.~B. 1969, \mnras, 145, 271

\bibitem[{{Levermore}(1984)}]{lev84}
{Levermore}, C.~D. 1984, Journal of Quantitative Spectroscopy and Radiative
  Transfer, 31, 149

\bibitem[{{Levermore} \& {Pomraning}(1981)}]{lp81}
{Levermore}, C.~D., \& {Pomraning}, G.~C. 1981, \apj, 248, 321

\bibitem[{{Li} {et~al.}(2011){Li}, {Krasnopolsky}, \& {Shang}}]{li11}
{Li}, Z.-Y., {Krasnopolsky}, R., \& {Shang}, H. 2011, \apj, 738, 180

\bibitem[{{Machida}(2008)}]{mcd08}
{Machida}, M.~N. 2008, \apjl, 682, L1

\bibitem[{{Machida} {et~al.}(2009){Machida}, {Omukai}, {Matsumoto}, \&
  {Inutsuka}}]{mcd09}
{Machida}, M.~N., {Omukai}, K., {Matsumoto}, T., \& {Inutsuka}, S.-I. 2009,
  \mnras, 399, 1255

\bibitem[{{Masunaga} \& {Inutsuka}(2000)}]{mi00}
{Masunaga}, H., \& {Inutsuka}, S.-i. 2000, \apj, 531, 350

\bibitem[{{Matsumoto} \& {Hanawa}(2003)}]{mh03}
{Matsumoto}, T., \& {Hanawa}, T. 2003, \apj, 583, 296

\bibitem[{{Mellon} \& {Li}(2008)}]{ml08}
{Mellon}, R.~R., \& {Li}, Z.-Y. 2008, \apj, 681, 1356

\bibitem[{{Miyoshi} \& {Kusano}(2005)}]{miyoshi}
{Miyoshi}, T., \& {Kusano}, K. 2005, Journal of Computational Physics, 208, 315

\bibitem[{{Myers} {et~al.}(2011){Myers}, {Krumholz}, {Klein}, \&
  {McKee}}]{myers11}
{Myers}, A.~T., {Krumholz}, M.~R., {Klein}, R.~I., \& {McKee}, C.~F. 2011,
  \apj, 735, 49

\bibitem[{{Omukai}(2000)}]{omukai00}
{Omukai}, K. 2000, \apj, 534, 809

\bibitem[{{Omukai} {et~al.}(2010){Omukai}, {Hosokawa}, \& {Yoshida}}]{ohy10}
{Omukai}, K., {Hosokawa}, T., \& {Yoshida}, N. 2010, \apj, 722, 1793

\bibitem[{{Omukai} {et~al.}(2005){Omukai}, {Tsuribe}, {Schneider}, \&
  {Ferrara}}]{omukai05}
{Omukai}, K., {Tsuribe}, T., {Schneider}, R., \& {Ferrara}, A. 2005, \apj, 626,
  627

\bibitem[{{Penston}(1969)}]{pen69}
{Penston}, M.~V. 1969, \mnras, 144, 425

\bibitem[{{Planck Collaboration} {et~al.}(2011){Planck Collaboration}, {Ade},
  {Aghanim}, {Arnaud}, {Ashdown}, {Aumont}, {Baccigalupi}, {Balbi}, {Banday},
  {Barreiro}, {Bartlett}, {Battaner}, {Benabed}, {Beno{\^i}t}, {Bernard},
  {Bersanelli}, {Bhatia}, {Bock}, {Bonaldi}, {Bond}, {Borrill}, {Bot},
  {Bouchet}, {Boulanger}, {Bucher}, {Burigana}, {Cabella}, {Cardoso},
  {Catalano}, {Cay{\'o}n}, {Challinor}, {Chamballu}, {Chiang}, {Chiang},
  {Christensen}, {Clements}, {Colombi}, {Couchot}, {Coulais}, {Crill},
  {Cuttaia}, {Danese}, {Davies}, {Davis}, {de Bernardis}, {de Gasperis}, {de
  Rosa}, {de Zotti}, {Delabrouille}, {Delouis}, {D{\'e}sert}, {Dickinson},
  {Dobashi}, {Donzelli}, {Dor{\'e}}, {D{\"o}rl}, {Douspis}, {Dupac},
  {Efstathiou}, {En{\ss}lin}, {Finelli}, {Forni}, {Frailis}, {Franceschi},
  {Fukui}, {Galeotta}, {Ganga}, {Giard}, {Giardino}, {Giraud-H{\'e}raud},
  {Gonz{\'a}lez-Nuevo}, {G{\'o}rski}, {Gratton}, {Gregorio}, {Gruppuso},
  {Harrison}, {Helou}, {Henrot-Versill{\'e}}, {Herranz}, {Hildebrandt},
  {Hivon}, {Hobson}, {Holmes}, {Hovest}, {Hoyland}, {Huffenberger}, {Jaffe},
  {Jones}, {Juvela}, {Kawamura}, {Keih{\"a}nen}, {Keskitalo}, {Kisner},
  {Kneissl}, {Knox}, {Kurki-Suonio}, {Lagache}, {L{\"a}hteenm{\"a}ki},
  {Lamarre}, {Lasenby}, {Laureijs}, {Lawrence}, {Leach}, {Leonardi}, {Leroy},
  {Linden-V{\o}rnle}, {L{\'o}pez-Caniego}, {Lubin}, {Mac{\'{\i}}as-P{\'e}rez},
  {MacTavish}, {Madden}, {Maffei}, {Mandolesi}, {Mann}, {Maris},
  {Mart{\'{\i}}nez-Gonz{\'a}lez}, {Masi}, {Matarrese}, {Matthai}, {Mazzotta},
  {Meinhold}, {Melchiorri}, {Mendes}, {Mennella}, {Miville-Desch{\^e}nes},
  {Moneti}, {Montier}, {Morgante}, {Mortlock}, {Munshi}, {Murphy}, {Naselsky},
  {Nati}, {Natoli}, {Netterfield}, {N{\o}rgaard-Nielsen}, {Noviello},
  {Novikov}, {Novikov}, {Onishi}, {Osborne}, {Pajot}, {Paladini}, {Paradis},
  {Pasian}, {Patanchon}, {Perdereau}, {Perotto}, {Perrotta}, {Piacentini},
  {Piat}, {Plaszczynski}, {Pointecouteau}, {Polenta}, {Ponthieu}, {Poutanen},
  {Pr{\'e}zeau}, {Prunet}, {Puget}, {Reach}, {Rebolo}, {Reinecke}, {Renault},
  {Ricciardi}, {Riller}, {Ristorcelli}, {Rocha}, {Rosset}, {Rowan-Robinson},
  {Rubi{\~n}o-Mart{\'{\i}}n}, {Rusholme}, {Sandri}, {Savini}, {Scott},
  {Seiffert}, {Smoot}, {Starck}, {Stivoli}, {Stolyarov}, {Sudiwala}, {Sygnet},
  {Tauber}, {Terenzi}, {Toffolatti}, {Tomasi}, {Torre}, {Tristram}, {Tuovinen},
  {Umana}, {Valenziano}, {Varis}, {Vielva}, {Villa}, {Vittorio}, {Wade},
  {Wandelt}, {Wilkinson}, {Ysard}, {Yvon}, {Zacchei}, \& {Zonca}}]{pldust}
{Planck Collaboration} {et~al.} 2011, \aap, 536, A17

\bibitem[{{Schneider} {et~al.}(2002){Schneider}, {Ferrara}, {Natarajan}, \&
  {Omukai}}]{schn02}
{Schneider}, R., {Ferrara}, A., {Natarajan}, P., \& {Omukai}, K. 2002, \apj,
  571, 30

\bibitem[{{Sch{\"o}nke} \& {Tscharnuter}(2011)}]{sch11}
{Sch{\"o}nke}, J., \& {Tscharnuter}, W.~M. 2011, \aap, 526, A139

\bibitem[{{Seaton} {et~al.}(1994){Seaton}, {Yan}, {Mihalas}, \&
  {Pradhan}}]{op94}
{Seaton}, M.~J., {Yan}, Y., {Mihalas}, D., \& {Pradhan}, A.~K. 1994, \mnras,
  266, 805

\bibitem[{{Semenov} {et~al.}(2003){Semenov}, {Henning}, {Helling}, {Ilgner}, \&
  {Sedlmayr}}]{semenov}
{Semenov}, D., {Henning}, T., {Helling}, C., {Ilgner}, M., \& {Sedlmayr}, E.
  2003, \aap, 410, 611

\bibitem[{{Tomida} {et~al.}(2010{\natexlab{a}}){Tomida}, {Machida}, {Saigo},
  {Tomisaka}, \& {Matsumoto}}]{tomida10b}
{Tomida}, K., {Machida}, M.~N., {Saigo}, K., {Tomisaka}, K., \& {Matsumoto}, T.
  2010{\natexlab{a}}, \apjl, 725, L239

\bibitem[{{Tomida} {et~al.}(2013){Tomida}, {Tomisaka}, {Matsumoto}, {Hori},
  {Okuzumi}, {Machida}, \& {Saigo}}]{tomida13}
{Tomida}, K., {Tomisaka}, K., {Matsumoto}, T., {Hori}, Y., {Okuzumi}, S.,
  {Machida}, M.~N., \& {Saigo}, K. 2013, \apj, 763, 6

\bibitem[{{Tomida} {et~al.}(2010{\natexlab{b}}){Tomida}, {Tomisaka},
  {Matsumoto}, {Ohsuga}, {Machida}, \& {Saigo}}]{tomida10a}
{Tomida}, K., {Tomisaka}, K., {Matsumoto}, T., {Ohsuga}, K., {Machida}, M.~N.,
  \& {Saigo}, K. 2010{\natexlab{b}}, \apjl, 714, L58

\bibitem[{{Tomisaka}(1998)}]{tmsk98}
{Tomisaka}, K. 1998, \apjl, 502, L163+

\bibitem[{{Tomisaka}(2002)}]{tmsk02}
---. 2002, \apj, 575, 306

\bibitem[{{Vaytet} {et~al.}(2013){Vaytet}, {Chabrier}, {Audit}, {Commer{\c
  c}on}, {Masson}, {Ferguson}, \& {Delahaye}}]{vaytet13}
{Vaytet}, N., {Chabrier}, G., {Audit}, E., {Commer{\c c}on}, B., {Masson}, J.,
  {Ferguson}, J., \& {Delahaye}, F. 2013, \aap, 557, A90

\bibitem[{{Winkler} \& {Newman}(1980)}]{wn80b}
{Winkler}, K.-H.~A., \& {Newman}, M.~J. 1980, \apj, 238, 311

\end{thebibliography}
\end{document}